\documentclass[11pt,twoside,onecolumn]{article}
\usepackage{graphicx,epsf}
\usepackage[]{latexsym,amsmath,amssymb}
\newcommand{\be}{\begin{eqnarray}}
\newcommand{\ee}{\end{eqnarray}}

\begin{document}
\title{\bf On brane-world black holes and short scale physics}
\author{Roberto Casadio\thanks{E-mail: casadio@bo.infn.it}
\\
\null
\\
{\em Dipartimento di Fisica, Universit\`a di
Bologna and I.N.F.N., Sezione di Bologna,}
\\
{\em via Irnerio 46, 40126 Bologna, Italy}}
\maketitle
\begin{abstract}
There is evidence that trans-Planckian physics does not affect
the Hawking radiation in four dimensions and, consequently,
deviations from the linear dispersion relation (for massless
particles) at very high energies cannot be revealed using
four-dimensional black holes.
We study this issue in the context of models with extra spatial
dimensions and show that small black holes that could be produced
in accelerators might also provide a chance of testing the high
energy regime where non-linear dispersion relations are generally
expected.
\end{abstract}
%
%\pacs{04.70.-s, 04.70.Bw, 04.50.+h}
%
\raggedbottom
\section{Introduction}
\setcounter{equation}{0}
\label{intro}
In recent years some progress has been achieved toward a better
understanding of the role played by short distance physics in the
semiclassical treatment of gravitational systems.
Trans-Planckian modes show up in quantum field theory
on the curved background of a black hole due to the Hawking
radiation \cite{hawking}.
The spectrum of the radiated quanta measured far away from
the horizon $r_h=2\,M$ of a Schwarzschild black hole is
expected to be Planckian with a typical energy $\omega$ of
the order of the Hawking temperature,
\be
\omega\sim k_B\,T_H=\left(8\,\pi\,M\right)^{-1}
\ ,
\label{flux}
\ee
where $M=(\ell_p/m_p)\,\bar M$ is the Arnowitt Deser Misner (ADM) mass
parameter of the Schwarzschild black hole ($k_B$ is Boltzmann constant
and we use units with $c=\hbar=1$, $m_p=\ell_p^{-1}$ is the Planck mass).
However, such (usually very soft) quanta were unboundedly energetic
if they originated at a radius $r_s$ very close to the horizon.
In fact, neglecting the backreaction, the gravitational blue-shift
is given by
\be
z=\sqrt{r_s\over r_s-r_h}-1
\ ,
\ee
and, in the geometrical optics approximation, the energy of
Hawking quanta at the point of emission
\be
\omega_s=(1+z)\,\omega
\ ,
\label{bs}
\ee
diverges for $r_s\to r_h$.
\par
Numerical studies of wave-packets moving toward the black hole have
shown that any modification of the radiation spectrum at
energies around the Planck mass $m_p$ or above do
not change the spectrum of the asymptotic flux \cite{unruh,jacobson}.
This could mean that either quanta originated at trans-Planckian
frequencies do not contribute or they are not excited at all.
Investigations based on the statistical mechanics of the
Hawking radiation (in the microcanonical ensemble) \cite{cr} and on
the uncertainty principle applied to the location of the source of
the radiation \cite{cmers} have further supported this conclusion.
All such investigations were carried out in four space-time
dimensions, where the fact that $\bar M$ {\em must\/} be larger than
$m_p$ in order to have a black hole plays a key role.
\par
Models with extra spatial dimensions have raised some interest
as possible scenarios in String Theory.
Their novel feature is that Standard Model fields are degrees of
freedom of open strings confined on a (negligibly thin)
four-dimensional submanifold (the so called {\em brane-world\/}),
while gravity is mediated by closed strings which propagate
in the whole space-time (the {\em bulk\/}).
Such models basically fall into two classes:
{\em i\/}) compact (flat) extra dimensions (the ADD scenario of
Refs.~\cite{add}) and
{\em ii\/}) (possibly infinite) warped extra dimensions
(the RS scenario of Refs.~\cite{RS}).
\par
The understanding of black holes in such models is still rather
incomplete to date (a partial list of References is given by
\cite{argyres,chamblin,emparan,ch,chp,maartens,cfm,wiltshire,kofinas,
kanti,katz,shiro,wiseman,cm,suranyi,fabbri}).
It is however clear that, since one of the main consequences of
the existence of extra dimensions is that the fundamental scale
of gravity $m_g$ ($=\ell_g^{-1}$) can be much smaller than $m_p$
(perhaps as small as a few TeV's), black holes might exist with proper
mass~\footnote[1]{The relation between the proper mass and the ADM
mass of a very small black hole is a subtle issue \cite{ch,chp,cfm,cm}.
We shall presently trust Newtonian reasoning and assume
they are equal \cite{argyres,emparan,katz}.}
$m_g<\bar M\ll m_p$ that could be produced in accelerators or
by cosmic rays (see, e.g.,
Refs.~\cite{banks,giddings,thomas,dimopoulos,ch1,ch2,tu,cavaglia,
cheung,kanti2,hofmann}).
The Hawking temperature for such black holes could be of the order
of $m_g$, the value at which non-standard dispersion relations
\cite{unruh,jacobson,laura}
might already be needed to effectively describe the physics by means
of quantum field theory (for a description of trans-Planckian
physics within the framework of String Theory, see,
e.g.~Refs.~\cite{tp}).
Deviations from the behaviour corresponding to a purely linear
dispersion relation in the Hawking spectrum at energies of the order
of a few TeV, which are within the range of planned experiments, might
therefore signal the existence of extra spatial dimensions.
\par
In these notes we intend to analyze such a possibility by studying
in a rather general framework the problem of Hawking radiation in
$4+d$ dimensions.
In Section~\ref{waves} we shall review why trans-Planckian modes are
not generated in four dimensions by appealing to the non-adiabaticity
of the time evolution of wave modes on a Schwarzschild background.
By repeating essentially the same argument in higher dimensional
space-times, we shall then show that for $d>0$ black holes with a
mass close to $m_g$ can emit radiation with energy of that order of
magnitude.
In Section~\ref{nonl} we explicitly consider the smooth dispersion
relation introduced in Ref.~\cite{laura} and show that its main effect
is to reduce the Hawking flux near $m_g$.
According to recent estimates \cite{dimopoulos} (see, however, the
remarks in Ref.~\cite{ch1}), such small black holes are expected to
decay instantaneously and the suppression of high energy modes
should therefore appear in the spectrum of the decay products.
\section{Scalar waves in brane-world black holes}
\setcounter{equation}{0}
\label{waves}
A (minimally coupled) scalar wave in $4+d$ dimensions
satisfies the equation ($\mu,\nu=0,\ldots,3+d$)
\be
\partial_\mu\,\left(
\sqrt{-g}\,g^{\mu\nu}\,\partial_\nu\right)
\Phi=0
\ ,
\label{kg}
\ee
where, for the case of interest, the metric tensor $g_{\mu\nu}$
(whose determinant is $g$) represents a black hole in the
brane-world.
\subsection{ADD scenario}
In the ADD scenario, the brane tension is neglected.
Hence, at least in principle, one should just supply the
$4+d$-dimensional vacuum Einstein equations with suitable
boundary conditions at the edges of the extra dimensions
\cite{ch,chp}.
However, this task proves already difficult enough and one
considers approximate (asymptotic) forms.
One usually writes the metric as
\be
ds^2=g_{tt}\,dt^2+g_{rr}\,dr^2+r^2\,d\Omega_{(2+d)}^2
\ ,
\ee
where $g_{tt}$ and $g_{rr}$ are the non-trivial metric elements
and $d\Omega_{(2+d)}$ the line element of a unit $2+d$-sphere.
The $3+d$-dimensional radial coordinate $r$ can be decomposed as
$r^2=r_b^2+y^2$, where $r_b$ is a three-dimensional radial
coordinate on the brane and $y$ is bounded from above by the
characteristic size $L$ of the extra dimensions.
Let us also denote by $r=r_h$ the horizon radius (this quantity will
in general depend on off-brane angles, since the shape of the
horizon cannot be exactly a sphere \cite{ch}).
\par
Outside the horizon of a large black hole, $L\ll r_h<r$, we can
neglect the extra dimensions.
One then simply takes the standard Schwarzschild line element on
the brane \cite{argyres},
\be
ds^2\simeq-\Delta_{(0)}(r_b)\,dt^2+{dr^2_b\over \Delta_{(0)}(r_b)}
+r^2_b\left(d\theta^2+\sin^2\theta\,d\phi^2\right)
+\sum_{i=1}^d\,dy_i^2
\ ,
\ee
where
\be
\Delta_{(d)}(x)=1-\left({r_h\over x}\right)^{1+d}
\ .
\ee
This is in fact a {\em black string\/}.
\par
For small black holes and $r_h<r\ll L$ \cite{argyres}, one can
analogously consider a spherically symmetric black hole in $4+d$
dimensions \cite{myers}.
However, in order to take into account the fact that $d$ spatial
dimensions have size $L$, we shall instead use the following
approximate form \cite{ch}
\be
ds^2\simeq-\Delta_{(d)}(r)\,dt^2+{dr^2\over \Delta_{(d)}(r)}+
r^2\,\left(d\theta^2+\sin^2\theta\,d\phi^2\right)
+r^2\,\sum_{i=1}^d\,d\phi_i^2
\ ,
\label{g4d}
\ee
where the $\phi_i$'s are off-brane angles such that
$dy_i\simeq r\,d\phi_i$.
\subsection{RS scenario}
In the RS scenario there is one extra dimension but the brane,
located at $y=0$, has a tension $\sigma$ which warps the bulk,
\be
ds^2=e^{-\sigma\,|y|}\,g_{ij}\,dx^i\,dx^j+dy^2
\ ,
\ee
where $g_{ij}=g_{ij}(x^i,y)$ reduces to flat four-dimensional
Minkowski in the absence of sources other than the brane itself
\cite{RS} and $\sigma^{-1}$ (roughly) plays the same role as $L$
in the ADD case.
\par
For large black holes (i.e.~$M\,\sigma\gg 1$), the horizon is flattened
on the brane in a pancake shape \cite{katz,cm} and the brane metric
departs but slightly from the four-dimensional Schwarzschild form.
Small black holes are instead believed to be closer in shape to
the corresponding ADD case, so that in their vicinity the metric
can be approximated as in Eq.~(\ref{g4d}) with $d=1$
\cite{katz,suranyi} (note that the warp factor $e^{-\sigma\,|y|}$
is essentially constant for $y\ll\sigma^{-1}$).
\subsection{Origin of Hawking radiation}
We are now ready to study the solutions of Eq.~(\ref{kg}) in
some generality.
Let us then assume that the scalar field $\Phi$ can be factorized
according to
\be
\Phi=e^{i\,\omega\,t}\,
R(r)\,S(\theta)\,e^{i\,m\,\phi}\,e^{i\,\sum\,n_i\,\phi_i}
\ ,
\label{phi_p}
\ee
with the $n_i$'s integers, so that $\Phi$ is periodic in the
off-brane angles.
Of course, such a field is {\em not\/} confined on the brane and
cannot therefore truly represent matter fields but can be used
to analyze gravitational waves.
In order to adjust for matter fields, one should replace the
last exponential above with a product of Dirac $\delta(\phi_i)$ or
some smooth realization of it (for more details, see~\ref{conf}).
That would just affect the coefficient $A$ which appears below,
and which we shall usually set to zero.
The radial equation obtained from Eq.~(\ref{kg}) for the metric
(\ref{g4d}) then becomes
\be
\left[-{\Delta_{(d)}\over r^{2+d}}\,{d\over dr}
\left(r^{2+d}\,\Delta_{(d)}\,{d\over dr}\right)
+\omega^2
-{\Delta_{(d)}\over r^2}\,A\right]R=0
\label{radial}
\ee
and the angular equation is
\be
\left[{1\over\sin\theta}\,{\partial\over\partial\theta}
\left(\sin\theta\,{\partial\over\partial\theta}\right)
-{m^2\over\sin^2\theta}\right]S=\lambda\,S
\ ,
\ee
where $A=\lambda+\sum\,n_i^2$ (unless confinement is enforced,
see~\ref{conf}), the separation constant $\lambda=l\,(l+1)$
and $S=Y_l^m$ is a standard three-dimensional spherical harmonic.
\par
The radial equation can be further simplified by defining the
``tortoise'' coordinate
\be
dr_*\equiv \Delta_{(d)}^{-1}\,dr
\ ,
\ee
and introducing a rescaled radial function,
\be
W\equiv r^{1+d/2}\,R
\ ,
\ee
which then satisfies a Schr\"odinger-like equation
\be
\left[-{d^2\over dr_*^2}+V_d\right]\,W=\omega^2\,W
\ ,
\ee
where the potential $V_d$ is given by
\be
V_d=\left[(1+d)\,\left(1+{d\over 2}\right)+ A\right]\,
{\Delta_{(d)}\over{r^{2}}}
-\,\left(1+{d\over 2}\right)^2\,{\Delta_{(d)}^2\over r^2}
\ .
\label{Vd}
\ee
This potential generates a barrier located outside
the horizon which suppresses the gray-body factor for all modes of
the scalar field including those with zero angular momentum.
The effect for $d=0$ is mild, however for $d>0$ the barrier increases,
significantly reducing the black hole decay rate \cite{chp}
(see also \ref{conf}).
\par
There is now a relatively easy way to understand the reason why
trans-Planckian modes do not play any role in four dimensions.
Consider a spherically symmetric wave-packet $\Psi$ built up with
in-falling modes which shrinks toward the center of a
Schwarzschild black hole from a very large radius.
Since for $r\gg r_h=2\,M$ the solutions to the radial equation
(\ref{radial}) can be approximated by spherical waves,
$R\sim R_k=e^{i\,k\,r}$, the packet can be written as
\be
\Psi(r,t)=
\int_0^\infty dk\,k^{2+d}\,f(k)\,e^{i\,\omega\,t}\,R_k
\simeq\int_0^\infty dk\,k^{2+d}\,f(k)\,e^{i\,\omega\,t+i\,k\,r}
\ ,
\label{packet}
\ee
where $\omega=k$ and $f(k)$ is, e.g., a Gaussian in momentum space.
It is actually impossible to follow the evolution of such a packet
analytically in time, however one can use numerical methods to show
that out-going modes start to be generated as the packet approaches
the horizon \cite{unruh}.
One can look at this effect in two equivalent ways, i.e.~as a violation
of the WKB approximation in a time-independent context or as a breakdown
of the adiabatic approximation in a time-dependent framework.
\par
For the latter viewpoint, consider that each mode in the sum
(\ref{packet}) moves in time according to the equation of motion
(\ref{kg}) and hence experiences the potential $V_d$.
According to the adiabatic approximation, such a potential does not
appreciably affect a mode $k$ if it does not change significantly
during the typical time $k^{-1}$ of oscillation of the mode.
During this time the mode itself will move a distance of order $k^{-1}$
toward the horizon, so that the adiabatic approximation is satisfied if
the difference $|V_d(r-k^{-1})-V_d(r)|$ is negligible.
\par
The above is also the condition for the validity of the WKB
approximation.
If one further considers the blue-shift (\ref{bs}) of the wave modes as
they approach toward the horizon, one finds the following bound
\be
\Sigma_d\equiv
\left|{V_d(r-\tilde k^{-1})-V_d(r+\tilde k^{-1})\over V_d(r)}\right|
\ll 1
\ ,
\ee
where $\tilde k^{-1}=\sqrt{\Delta_{(d)}}\,k^{-1}$ is the blue-shifted
wavelength.
The ratio $\Sigma_d$ can be easily computed, although its explicit
expression appears rather cumbersome and we omit it.
\subsubsection{Four-dimensional case}
\begin{figure}
\center{
\raisebox{3cm}{$\Sigma_0$}
\epsfxsize=2.9in
\epsfbox{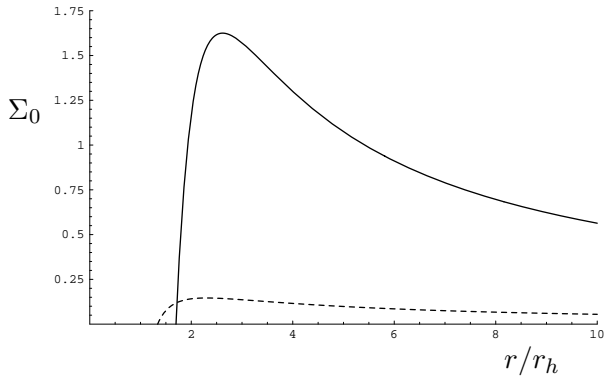}\\
\hspace{-0.2in}
\raisebox{0.5cm}
{\hspace{6.5cm}$r/r_h$}
}
\caption{The ratio $\Sigma_{d=0}$ in four dimensions for
$\lambda=r_h$ (solid line) and $0.1\,r_h$ (dashed line).}
\label{four}
\end{figure}

The quantity $\Sigma_0$ (for $s$-waves with $A=0$) as a function
of $r$ for the two wavelengths $k^{-1}=\lambda=r_h$ and $r_h/10$
is plotted in Fig.~\ref{four}.
For $\lambda=r_h$ the relative change in the potential becomes
of order $1$ already at $r\simeq 5\,r_h$.
This suggests that backscattered (out-going) modes really come
from a region of radius appreciably larger than the horizon,
that is $2\,r_h\lesssim r_s\lesssim10\,r_h$.
The corresponding blue-shift [defined in Eq.~(\ref{bs})] lies
therefore in the range $0.05\lesssim z\lesssim 0.41$.
Its average is in good quantitative agreement with the result of
Ref.~\cite{cmers}, where an effective blue-shift $z\simeq 0.23$
was obtained by a different averaging prescription over the
source position.
It is also clear from the plot that the modes with wavelength
much shorter than $r_h$ practically fall in a constant potential
(the relative change $\Sigma_0<0.15$ for $\lambda=r_h/10$
and it is smaller for even shorter $\lambda$) and are not
appreciably backscattered.
\par
Considering again the packet (\ref{packet}) one is thus led to
conclude that when it approaches the horizon, it is progressively
distorted, with in-falling components of wavelength $\lambda\sim r_h$
being (partially) backscattered, while higher frequency components
fall across the horizon undisturbed.
The same effect occurs to the vacuum surrounding a black hole:
virtual particles with wavelength of order $r_h$ are backscattered
by the potential barrier, while their peers tunnel through inside
the horizon;
higher frequency modes behave instead as in flat space.
\subsubsection{Extra-dimensional cases}
\begin{figure}
\center{
\raisebox{3cm}{$\Sigma_d$}
\epsfxsize=2.9in
\epsfbox{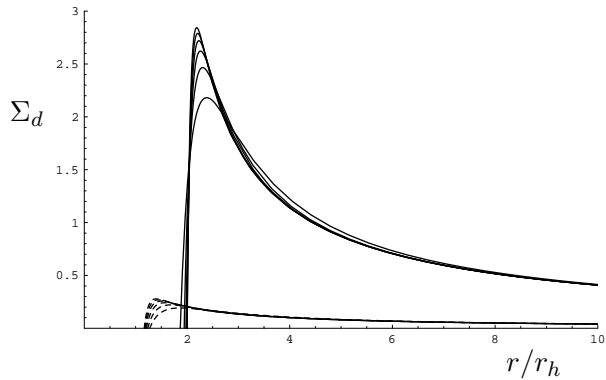}\\
\hspace{-0.2in}
\raisebox{0.5cm}
{\hspace{6.5cm}$r/r_h$}
}
\caption{The ratio $\Sigma_d$ for $\lambda=r_h$ (solid lines) and
$0.1\,r_h$ (dashed lines).
$d=1,\ldots,6$ (lower up).}
\label{extra}
\end{figure}

With increasing $d=1,\ldots,6$, the situation for $\lambda=r_h$
and $r_h/10$ is that of Fig.~\ref{extra}, which is qualitatively
the same as in four dimensions.
Let us also recall that $r_h$ in the ADD scenario depends on
$d$, roughly according to the expression \cite{argyres,ch,chp}
\be
r_h\simeq \left(2\,C_d\,L^d\,M\right)^{1\over 1+d}
\ ,
\label{R_H<}
\ee
where $C_d$ is a numerical coefficient which accounts for the
geometry of the extra dimensions in relating $m_g$ to $m_p$
\cite{add,argyres,myers}, and \cite{add}
\be
L\simeq \left[{1\,{\rm TeV}\over m_g}\right]\,
10^{{31\over d}-19}\,{\rm m}
\ee
is the size of the extra dimensions (the case $d=1$ is ruled out
since it would imply corrections to the Newton law at solar system
scale and below \cite{add}).
On assuming $m_g\sim 1\,$TeV, $C_d\sim 1$ and $M\sim 10^{-48}\,$m
($\bar M\simeq 10\,m_g$), one has the typical horizon radius (see
Table~\ref{t1})
\be
r_h\sim 10^{-19}\,{\rm m}\sim \ell_g
\ .
\label{rh}
\ee
Thus, $\bar M\sim 10\,m_g$ is the minimum allowed mass for a black hole
in this context.
Since $\Sigma_d(\lambda=r_h)\sim 1$ for $r\sim 5\,r_h$, and
irrespectively of the value of $d>0$, one concludes
that quanta with the typical Hawking energy
\be
\omega\sim k_B\,T_H=m_p\,{\ell_p\over L}\,
\left({\ell_p\over M}\right)^{1\over 1+d}
\sim 1\,{\rm TeV}
\ ,
\ee
are again originated quite far away from the horizon and their
blue-shift is negligibly small (see Table~\ref{t1}).
\par
In the RS scenario, one expects $L\sim\sigma^{-1}<10^{-3}\,$m
and for a $10\,$TeV black hole Eq.~(\ref{R_H<}) with $d=1$ yields
the same order of magnitude as in Eq.~(\ref{rh}) for the ADD case.
\begin{table}
\center{
\begin{tabular}{|c|c|c|c|c|c|c|}
\hline
$d$ & 1 & 2 & 3 & 4 & 5 & 6 \\
\hline
$r_h$
& $1\times 10^{-18}$
& $5\times 10^{-19}$
& $3\times 10^{-19}$
& $3\times 10^{-19}$
& $2\times 10^{-19}$
& $2\times 10^{-19}$
\\
\hline
$z$
& $2\times 10^{-2}$
& $4\times 10^{-3}$
& $8\times 10^{-4}$
& $2\times 10^{-4}$
& $3\times 10^{-5}$
& $6\times 10^{-6}$
\\
\hline
${\mathcal L}_1$
& $1.0$
& $0.99$
& $0.98$
& $0.98$
& $0.96$
& $0.96$
\\
\hline
${\mathcal L}_{1/2}$
& $0.99$
& $0.98$
& $0.93$
& $0.93$
& $0.86$
& $0.86$
\\
\hline
\end{tabular}
\caption{Horizon radius $r_h$ (in meters) and blue-shift $z$
for modes with $k^{-1}\sim\ell_g$ of a $10\,$TeV black hole
with $d$ extra dimensions.
The luminosity per particle specie ${\mathcal L}_\alpha$ is
computed using the Epstein dispersion relation (\ref{eps}) and
expressed in units of the corresponding four-dimensional
canonical luminosity for $\alpha=1$ and $1/2$.
The case with $d=1$ in the ADD scenario is ruled out by
measurements \cite{add}.}
\label{t1}
}
\end{table}
\section{Non-linear dispersion relations}
\setcounter{equation}{0}
\label{nonl}
Since we have found that black holes can be used to produce
quanta with energy close to the fundamental scale $m_g\sim 1\,$TeV,
we shall now study how modified dispersion relations for such
quanta affect the evaporation.
We shall basically follow what was done in Ref.~\cite{cr}
and replace $m_p$ with the much smaller $m_g$.
\par
In Ref.~\cite{cr} we showed that deviations from linearity in the
spectrum yield a modified occupation number density for the
Hawking quanta with energy close to the fundamental scale.
The relation between the occupation number and the dispersion
relation is given by \cite{cr,visser}
\be
n(\omega)={1\over e^{4\,\pi\,r_h\,k_{\rm out}(\omega)}-1}
\ .
\label{n}
\ee
Note that the four-dimensional canonical number density
\cite{hawking} is formally recovered for the linear dispersion
relation $k_{\rm out}=\omega$.
\par
\begin{figure}
\center{
\raisebox{3cm}{$\omega$}
\epsfxsize=2.9in
\epsfbox{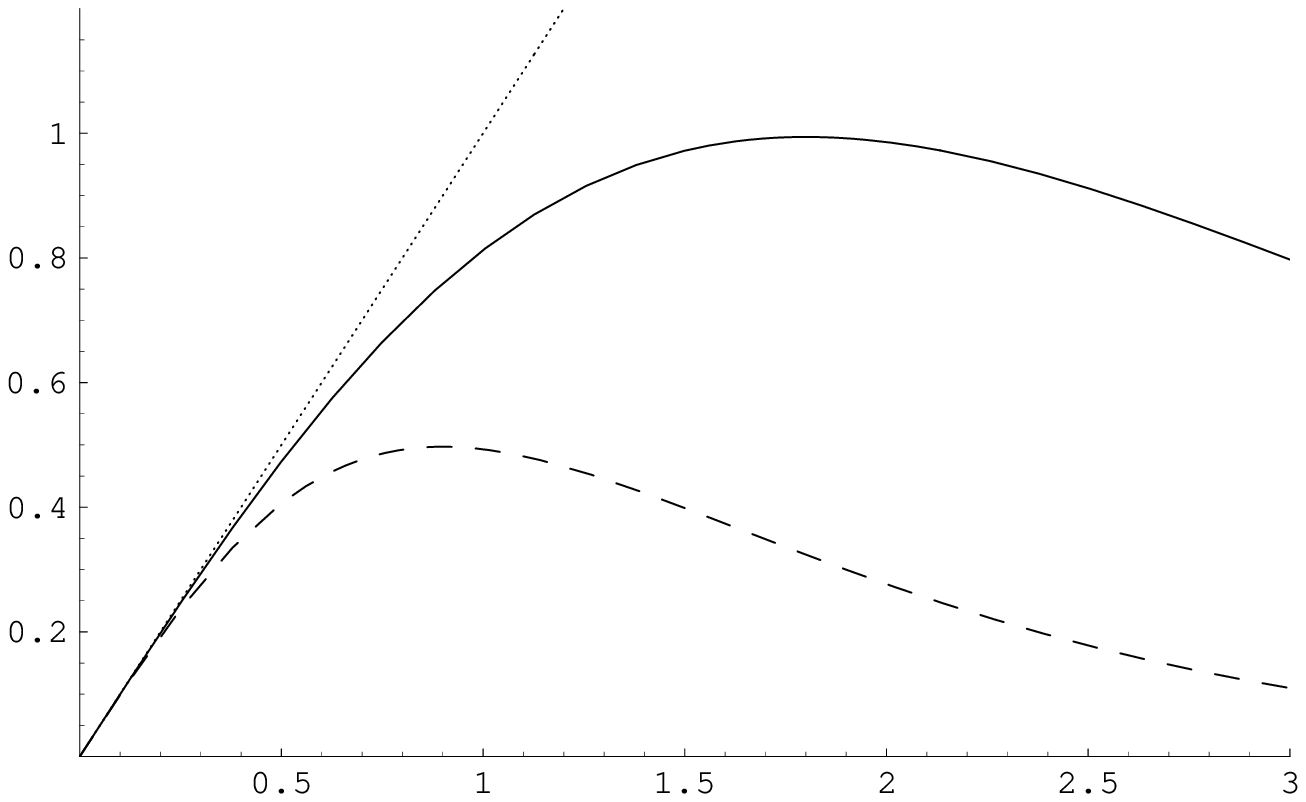}\\
\hspace{-0.2in}
\raisebox{0.5cm}
{\hspace{6.5cm}$k_{\rm out}$}
}
\caption{The spectrum of Eq.~(\ref{eps}) for $\alpha=1$
(solid line) and $\alpha=1/2$ (dashed line) compared
with the linear dispersion relation (dotted line).
The variables $\omega$ and $k_{\rm out}$ are in TeV.}
\label{epstein}
\end{figure}

As an example of dispersion relation we shall again take the
Epstein function \cite{laura} with a maximum at
$\omega\simeq \alpha\,m_g=\alpha\times 1\,$TeV,
\be
\omega=
k_{\rm out}\,
{\rm sech}\left({2\,k_{\rm out}\over3\,\alpha}\right)
\ ,
\label{eps}
\ee
which is plotted in Fig.~\ref{epstein} for $\alpha=1$ and $1/2$
(the parameter $\alpha$ accounts for our ignorance of the details
at high energy).
Note that, at this point, all the dependence on $d$ is contained
in $r_h$ as a function of the proper mass $\bar M$ as given
in Eq.~(\ref{R_H<}).
\par
From the number density the luminosity of the black hole
per particle specie follows straightforwardly,
\be
{\mathcal L}={\mathcal A}\,\int_0^\infty
\Gamma(\omega)\,n(\omega)\,\omega^3\,d\omega
\equiv
{\mathcal A}\,\int_0^\infty
\Gamma(\omega)\,E(\omega)\,d\omega
\ ,
\label{L}
\ee
where $\Gamma\sim 1$ is the gray-body factor for the given
particle specie and ${\mathcal A}=16\,\pi\,r_h^2$ the
four-dimensional horizon area.
This expression neglects gravitational waves emitted
into the bulk, since there is evidence that they contribute
negligibly to the overall luminosity, because there are far less
bulk graviton modes than Standard Model particles
\cite{emparan,ch} (see, however, Refs.~\cite{frolov}
for different pictures, and \ref{dphase} for an estimate of
the bulk emission).
One then gets
\be
{\mathcal L}_{(d)}(\bar M)\simeq
4\,\pi\,r_h^2(\bar M)\,\int_0^\infty
{\rm sech}^4\left({2\,k\over 3\,\alpha}\right)\,\left|
1-{2\,k\over 3\,\alpha}\,
\tanh\left({2\,k\over 3\,\alpha}\right)\right|\,
{k^3\,dk\over e^{4\,\pi\,r_h(\bar M)\,k}-1}
\ .
\label{lumi}
\ee
Note that, since $r_h>1\,$TeV$^{-1}$, the contribution from
$k>k_m$, ($k_m\simeq 1.8\,\alpha\,$TeV is the value of $k_{\rm out}$
which maximizes $\omega$), is negligible because of the
exponential suppression at large $k$ and could be omitted.
The luminosity can be integrated numerically and, for
$\bar M=10\,$TeV, one finds the values given in Table~\ref{t1},
where we display the ratio
\be
{\mathcal L}_\alpha\equiv {{\mathcal L}_{(d)}\over
{\mathcal L}_H}
\ ,
\ee
between the computed luminosity and the canonical luminosity
${\mathcal L}_H$ of a four-dimensional black hole of the same
mass \cite{hawking} and $\alpha=1$ and $1/2$.
For larger values of $\bar M$ (hence of $r_h$) the ratio between
the two quantities quickly approaches one (the same result was
previously obtained in four dimensions for $\bar M\ge m_p$
\cite{cr}).
\par
\begin{figure}
\center{
\raisebox{3cm}{$E$}
\epsfxsize=2.9in
\epsfbox{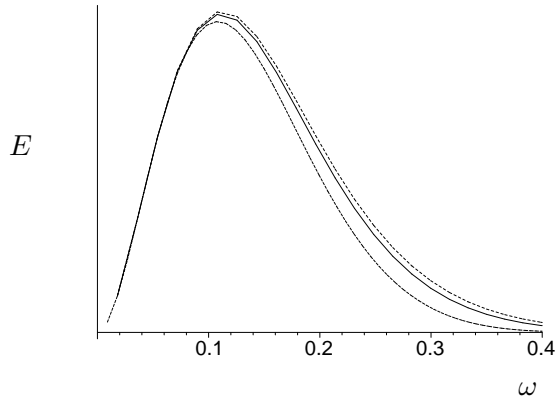}\\
\hspace{-0.2in}
\raisebox{0.5cm}
{\hspace{6.5cm}$\omega$}
}
\caption{The energy $E$ (divided by the horizon area) radiated in the
range of frequencies between $\omega$ and $\omega+d\omega$ by a black
hole in $4+d$ dimensions according to the dispersion relation (\ref{eps})
with $\alpha=1$ (solid line) and $\alpha=1/2$ (dashed line) and to
the linear dispersion relation (dotted line).
Vertical units are arbitrary.}
\label{plotE}
\end{figure}
\begin{figure}
\center{
\raisebox{3cm}{${E\over E_H}$}
\epsfxsize=2.9in
\epsfbox{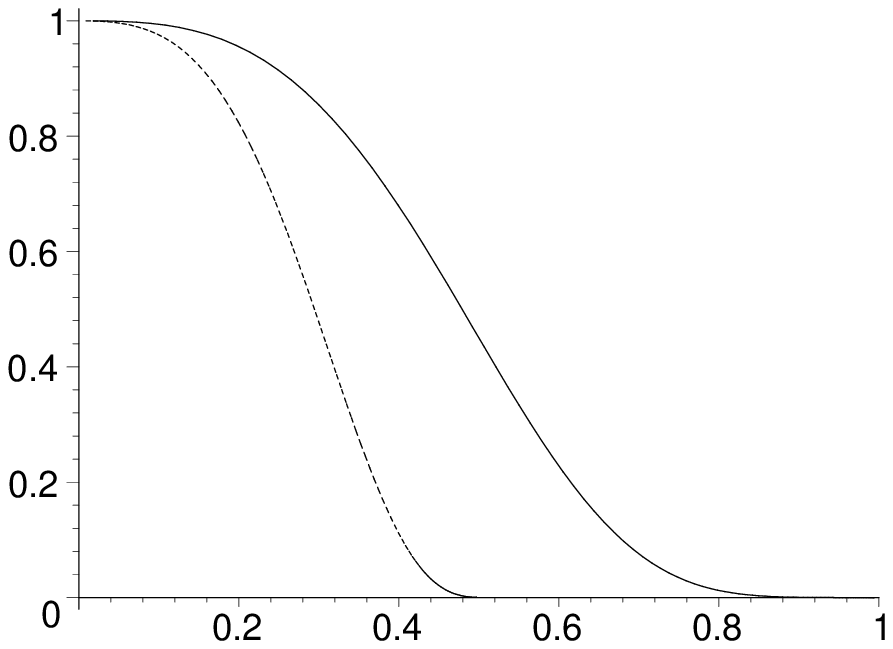}\\
\hspace{-0.2in}
\raisebox{0.5cm}
{\hspace{6.5cm}$\omega$}
}
\caption{The ratio between the energy radiated according to the
dispersion relation (\ref{eps}) with $\alpha=1$ (solid line)
and $\alpha=1/2$ (dashed line) and to the linear dispersion
relation.}
\label{plotEr}
\end{figure}

This result implies that, in order to test non-linear dispersion
relations, one should better look for exclusive (rather than inclusive)
effects, such as the probability of emitting particles with energy
in a given range (close to $m_g$) or the energy $E(\omega)$ emitted
in the range of frequencies between $\omega$ and $\omega+d\omega$.
The latter quantity is easily evaluated for the canonical number
density, we shall call it $E_H$, and can be used for comparison.
It can also be numerically determined for the dispersion relation
(\ref{eps}) and the result is plotted in Figs.~\ref{plotE} and
\ref{plotEr} for $\alpha=1$ and $1/2$.
If one trusts the picture according to which such small black holes
decay instantaneously \cite{dimopoulos}, then one expects an
energy spectrum for the decay products of the form in
Fig.~\ref{plotE}.
\section{Conclusions}
\setcounter{equation}{0}
\label{conc}
We have studied the high energy behaviour of Hawking radiation
in space-times with extra spatial dimensions in order to
highlight the possibility of using small black holes as
probes of non-linear dispersion relations.
Such deviations from linearity are in general expected within
the framework of String Theory and other theories of Quantum
Gravity, for which there is no certain experimental evidence
at present.
\par
The result of Section~\ref{waves}, that black holes in the brane-world
could emit particles with energy near the fundamental scale of gravity
$m_g$, is not surprising.
In fact, it simply means that the effect scales with $m_g$
naively:
as Planckian frequencies would be emitted in four dimensions
(where $m_g=m_p$)
by a Planck size black hole, energies of the order of $m_g$ are
generated by black holes with a mass close to $m_g$.
That is where non-linear dispersion relations should become
visible \cite{cr,cmers,tp}.
\par
In Section~\ref{nonl} we have explicitly analysed one of such
non-linear dispersion relations \cite{laura} and shown that
it induces a (relatively mild) suppression of high energy modes
which might be found in the decay products of small black
holes produced in accelerators.
Unfortunately, the exponential factor in the occupation number density
(\ref{n}) dominates over the modified dispersion relation for all
values of $d=0,\ldots,6$ and makes it hard to detect any effect
in the high energy regime ($\omega\sim m_g$).
\par
Let us conclude by remarking that the results we have obtained
concern objects which can be treated as classical black holes,
i.e.~whose horizon radius is larger than the Compton wavelength.
This condition is doomed to break down during the evaporation
process.
However, when the mass of the object has become sufficiently
small, it is still unlikely that trans-Planckian modes get
excited since there is not enough energy stored in the remnant.
At that point the canonical description obviously fails and
one should rely on the microcanonical description (see,
e.g.~\cite{micro} and References therein).
In four space-time dimensions, the latter approach was
shown in Ref.~\cite{cr} to change but slightly the overall
picture obtained for non-linear dispersion relations within
the canonical description and one does not expect a different
situation with extra spatial dimensions.
\appendix
\section{Confined fields}
\setcounter{equation}{0}
\label{conf}
As mentioned in Section~\ref{waves}, a field confined on the brane
cannot have a dependence on the extra spatial coordinates of
the form in Eq.~(\ref{phi_p}) and the separation constant $A$
must be correspondingly adjusted.
We shall here bring some evidence in favour of the fact that
the extra coordinate dependence does not appreciably affect the
results given in Sections~\ref{waves} and \ref{nonl} and that
setting $A=0$ is not a serious limitation.
\par
We focus on the dependence of the field $\Phi$ on the off-brane
angles $\phi_i$'s for  which we assume a smoothed out
$\delta(\phi_i)$ in the form of the Fourier sum
\be
\Phi\propto\prod_{i=1}^d\,{1\over\sqrt{2\,\pi}}\,
\sum_{n_i=-N}^N\,e^{i\,n_i\,\phi_i}
\ ,
\ee
where $N\gtrsim 2$ to ensure that $\Phi$ is peaked on the brane
($\phi_i\simeq 0$ for $i=1,\ldots,d$).
Then
\be
\partial_{\phi_i}^2\Phi\propto
-\prod_{i=1}^d\,{1\over\sqrt{2\,\pi}}\,
\sum_{n_i=-N}^N\,n_i^2\,e^{i\,n_i\,\phi_i}
\ .
\ee
Such a form clearly introduces a coupling between the different
$n_i$-modes which can be taken care of by integrating over
the angles $\phi_i$'s.
This introduces an effective separation constant
\be
A_N\simeq -\int_{-\pi}^{+\pi} \prod_{i=1}^d\,d\phi_i\,
\Phi\,\partial_{\phi_i}^2\Phi
=\left[{N\,(1+N)\,(1+2\,N)\over 6\,\pi}\right]^d
\ .
\ee
The numerical values are displayed in Table~\ref{t2} for
$N=2$ and $10$.
It appears that the more modes are included (i.e.~the larger is $N$)
and the larger is $d$, the more the corresponding $A_N$ dominates
over the contribution given in Eq.~(\ref{Vd}).
The overall effect is to increase enormously the potential
barrier which surrounds the horizon (thus suppressing the
Hawking emission), since the maximum of $V_d$ is roughly of the
same order of magnitude as $A_N$ for $A_N\gg 1$~\footnote[2]
{One cannot take this result too literally because of the
great simplifications we have assumed for the background
metric.
It is however suggestive to note that the more confined the
fields are, the weaker appears the Hawking radiation.}.
However, the ratios $\Sigma_d$ are just mildly affected.
For instance, the peak in the curve for $d=6$ as shown in
Fig.~\ref{extra} moves down from $2.9$ to $2.2$ for $A_{10}$
and $\Sigma_6$ is still roughly $1$ for $r\simeq 5\,r_h$.
To conclude, the qualitative analysis of Sections~\ref{waves}
and \ref{nonl} is not changed by the presence of $A\sim A_N\not=0$,
however large it is.

\begin{table}
\center{
\begin{tabular}{|c|c|c|c|c|c|c|}
\hline
$d$ & 1 & 2 & 3 & 4 & 5 & 6 \\
\hline
$A_2$
& $1.59$
& $2.53$
& $4.03$
& $6.41$
& $10.2$
& $16.3$
\\
\hline
$A_{10}$
& $123$
& $1.5\times 10^4$
& $1.8\times 10^6$
& $2.3\times 10^8$
& $2.8\times 10^{10}$
& $3.4\times 10^{12}$
\\
\hline
${\mathcal L}_1$
& $4.8\times 10^{-1}$
& $2.5\times 10^{-1}$
& $1.3\times 10^{-1}$
& $7.7\times 10^{-2}$
& $4.3\times 10^{-2}$
& $2.7\times 10^{-2}$
\\
\hline
\end{tabular}
\caption{The contribution to the separation constant $A_N$ from
the off-brane angle dependence for $N=2$ and $10$,
and the bulk luminosity per particle specie (in units of the
four-dimensional luminosity) for a black hole with
$\bar M=10\,$TeV and $\alpha=1$.
We recall that the case with $d=1$ in the ADD scenario is
ruled out by measurements \cite{add}.}
\label{t2}
}
\end{table}
\section{Bulk emission}
\setcounter{equation}{0}
\label{dphase}
The luminosity in Eqs.~(\ref{L}) and (\ref{lumi}) describes
the emission of four-dimensional particles (along the brane).
In order to take into account gravitational waves emitted into
the bulk, one must adjust both the horizon area
\be
{\mathcal A}_{(d)}\simeq
{2\,\pi^{3+d\over 2}\,r_h^{2+d}\over
\Gamma\left({3+d\over 2}\right)}
\ ,
\ee
and the phase space measure which becomes $d^{2+d}k$.
One then obtains the values given in Table~\ref{t2} for $\alpha=1$,
which, being smaller than the corresponding entries in Table~\ref{t1},
are negligible.
This is in agreement with the results of Refs.~\cite{emparan,ch},
according to which bulk gravitons do not contribute appreciably.
\end{document}